\documentclass[twoside]{article}
\usepackage{fleqn,espcrc2}

\usepackage{graphicx}
\usepackage[figuresright]{rotating}

\def\g {\gamma}

\newcommand{\AmS}{{\protect\the\textfont2
  A\kern-.1667em\lower.5ex\hbox{M}\kern-.125emS}}

\title{Pomeron intercept from BFKL gluon dynamics in
deep inelastic charm production at HERA}

\author{S.P.~Baranov\address[Lebedev]{P.N.~Lebedev Physics Institute,
    Russian Academy of Sciences,\\Leninsky prosp. 53, Moscow 117924, Russia},
     H.~Jung\address[Lund]{Particle Physics, Institute of
    Physics, Lund University, \\
        P.O. Box 118, 22100 Lund, Sweden} and
     N.P.~Zotov\address[Skobeltsyn]{D.V.~Skobeltsyn Institute of
    Nuclear Physics, M.V.~Lomonosov Moscow State University \\ 119899 Moscow,
Russia}
}
\begin{document}

\begin{abstract}
In the framework of semihard (k$_T$ factorization) QCD approach, we
consider the differential cross sections of $D^{*\pm}$ meson production at HERA.
The consideration is based on BFKL and CCFM gluon distributions. We find 
that in the case of BFKL LO gluon distribution the theoretical results
are sensitive to the Pomeron intercept parameter $\Delta$. We present
a comparison of the theoretical results with available ZEUS experimental
data.
\end{abstract}

\maketitle

\section{INTRODUCTION}

The experimental results on deep inelastic charm production obtained
 by the H1 \cite{1} and ZEUS \cite{2,3} Collaborations at HERA
provide a strong impetus for further theoretical studies. This 
process is truly semihard because of the presence of two larges scales:
the virtuality of the exchanged photon ($Q^2$) and the charm mass ($m_c^2$),
both being much larger than $\Lambda_{QCD}$ but much smaller than $s$.
Therefore, in the present note, we focus on the semihard approach~\cite{4,5}
(SHA), which we had applied earlier to the $D^{*\pm}$ meson 
photoproduction \cite{6,7} in a similar manner.

\section{THE SEMIHARD APPROACH}

\begin{figure*}[ht]
\begin{center}
\includegraphics[width=0.31\linewidth]{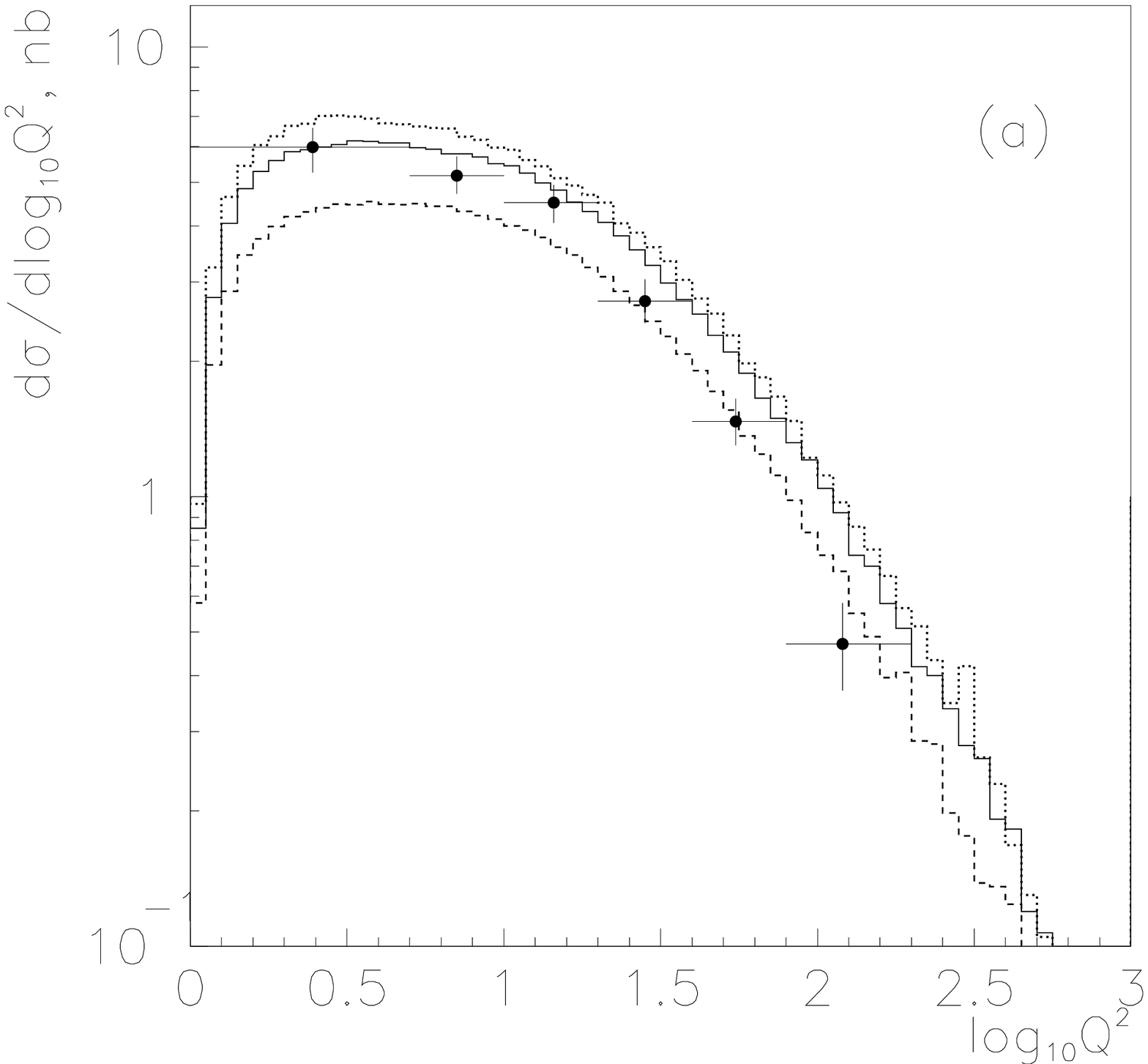}
\includegraphics[width=0.31\linewidth]{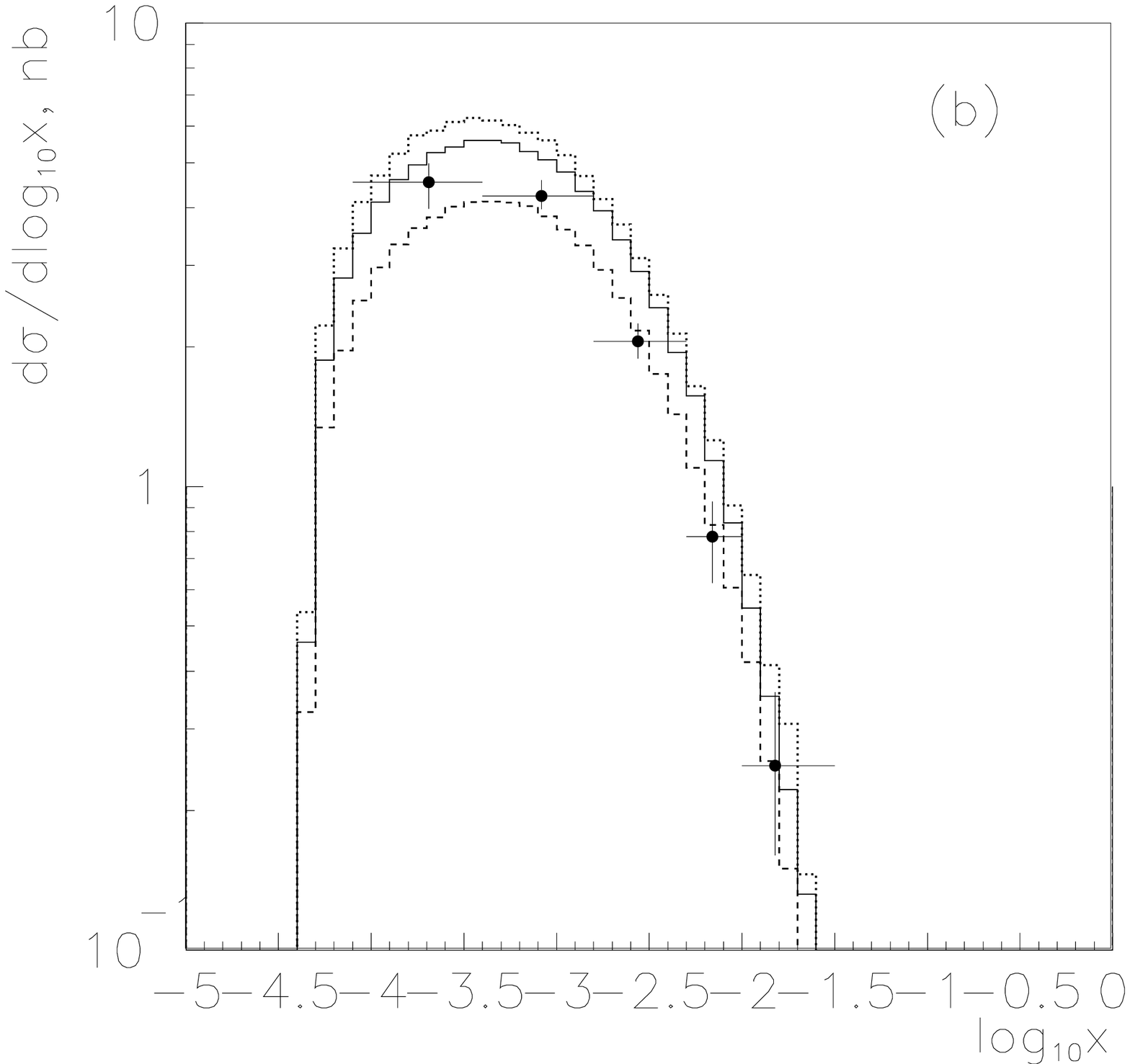}
\includegraphics[width=0.31\linewidth]{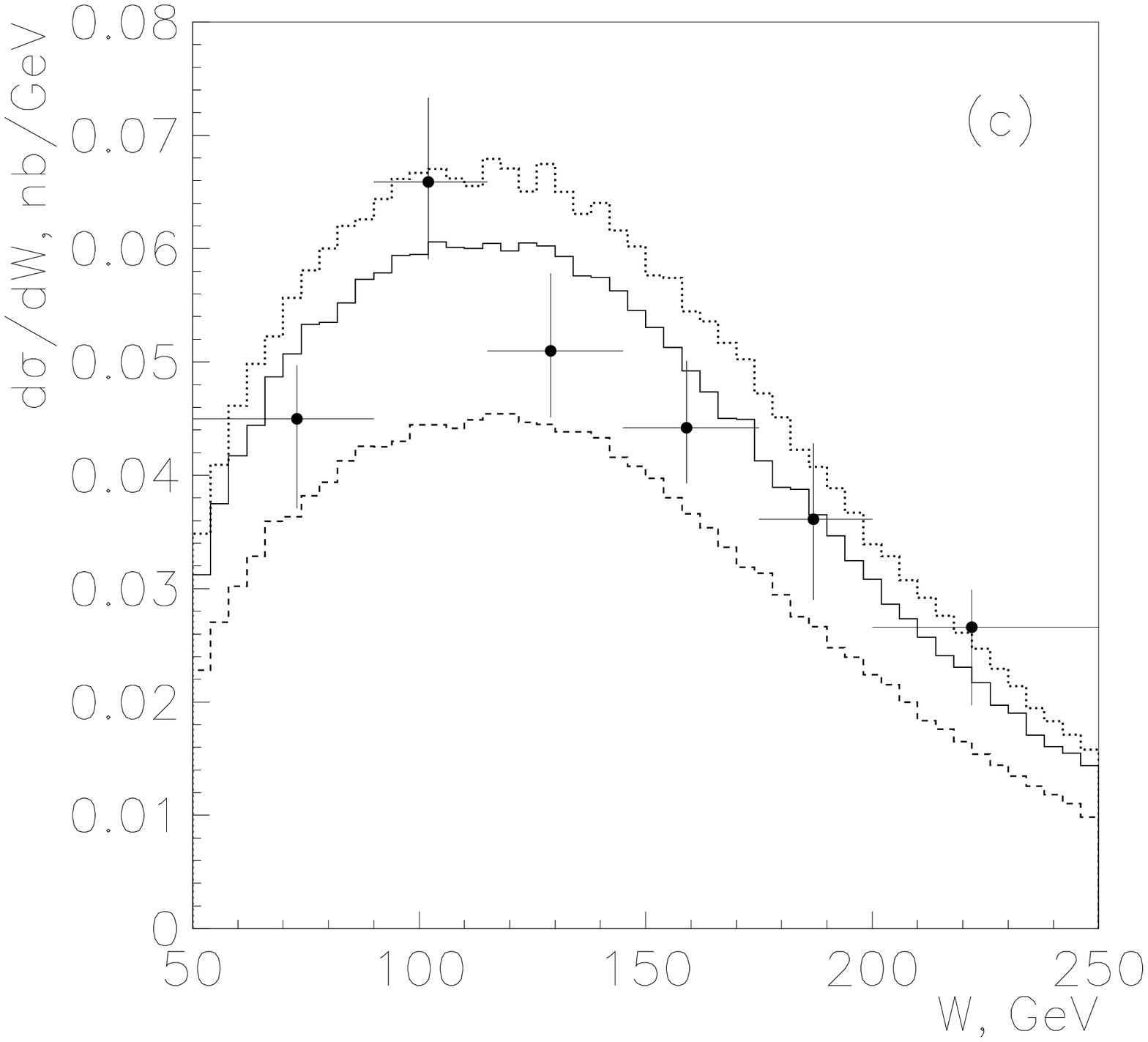}
\includegraphics[width=0.31\linewidth]{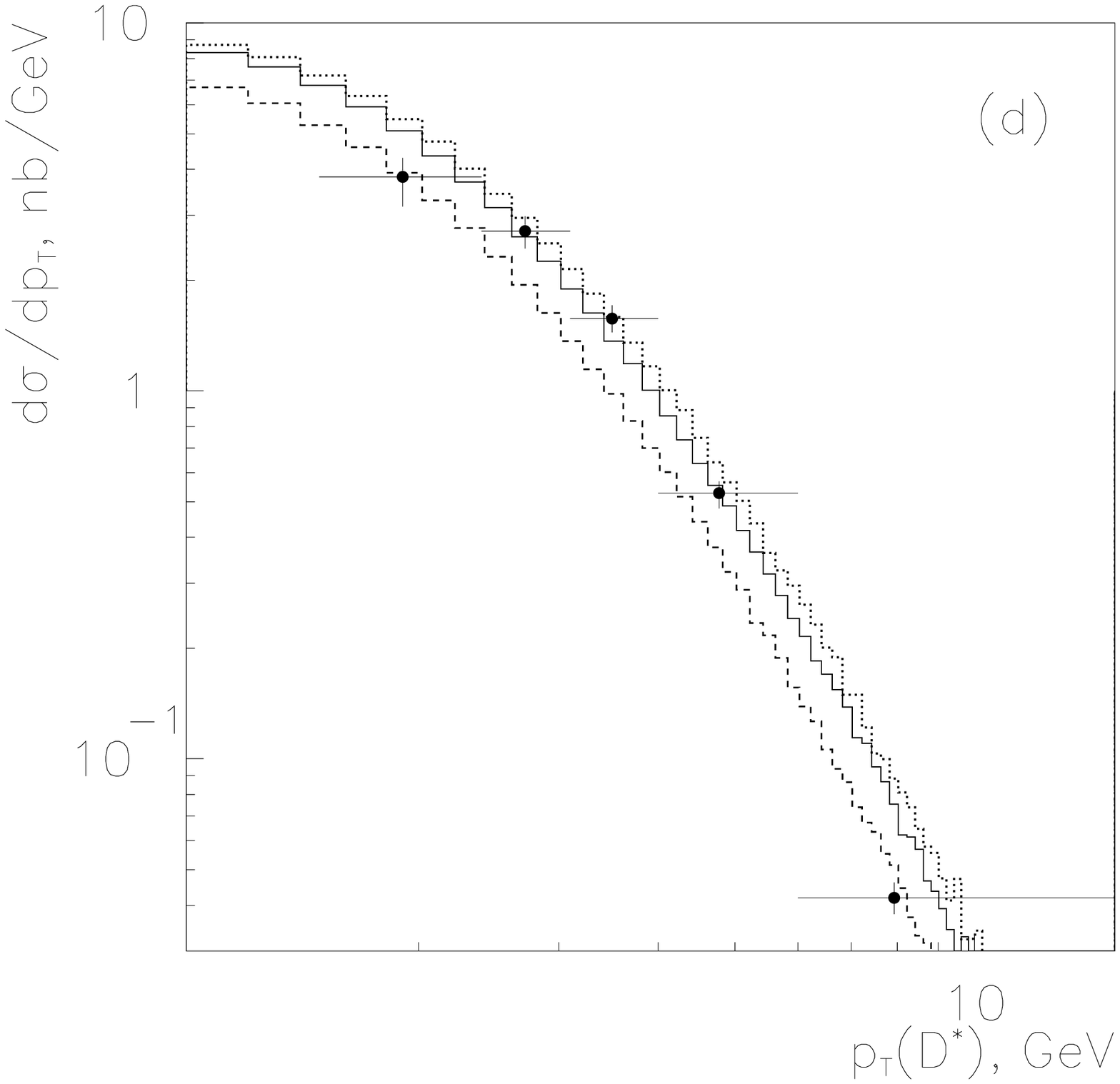}
\includegraphics[width=0.31\linewidth]{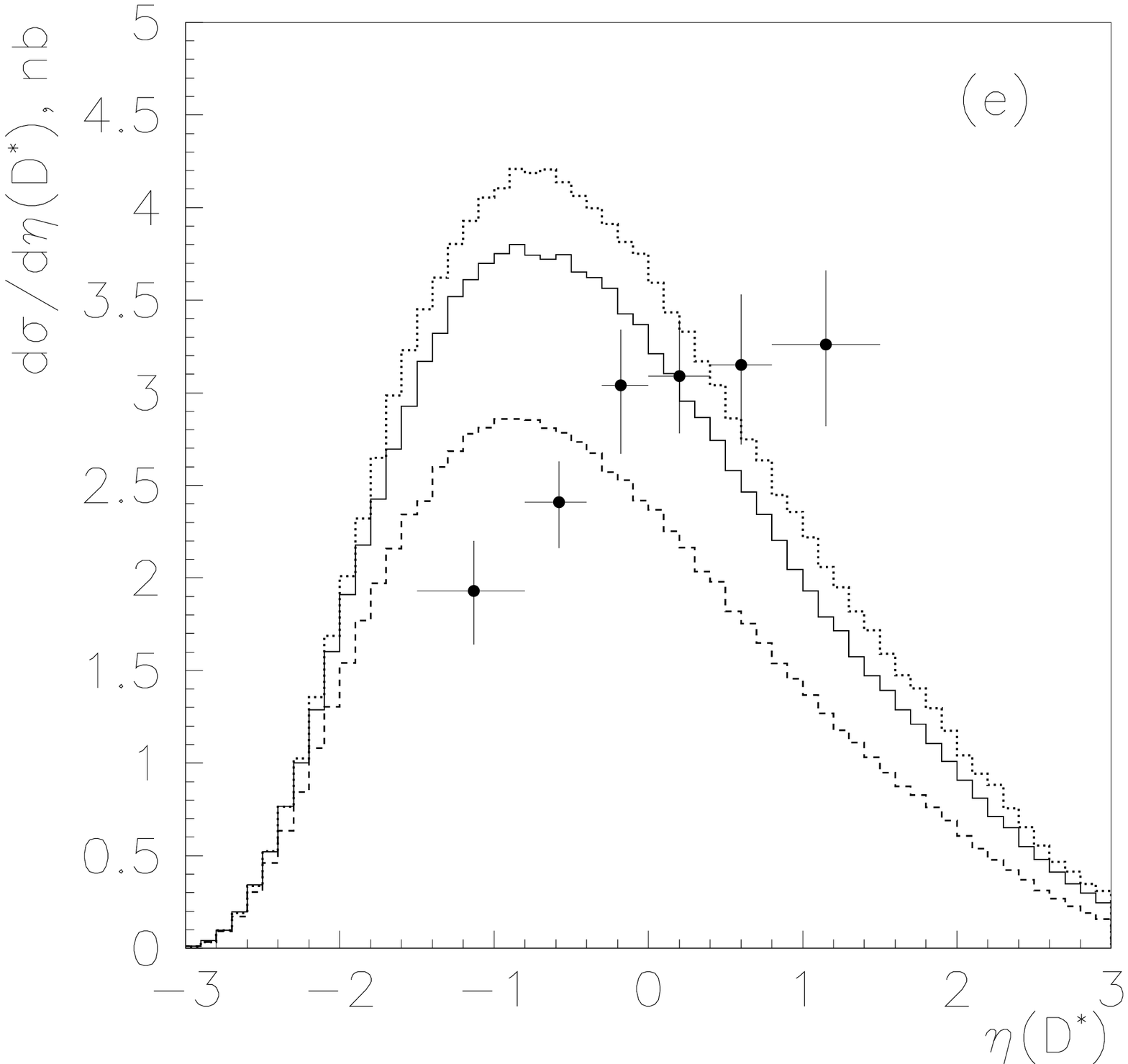}
\includegraphics[width=0.31\linewidth]{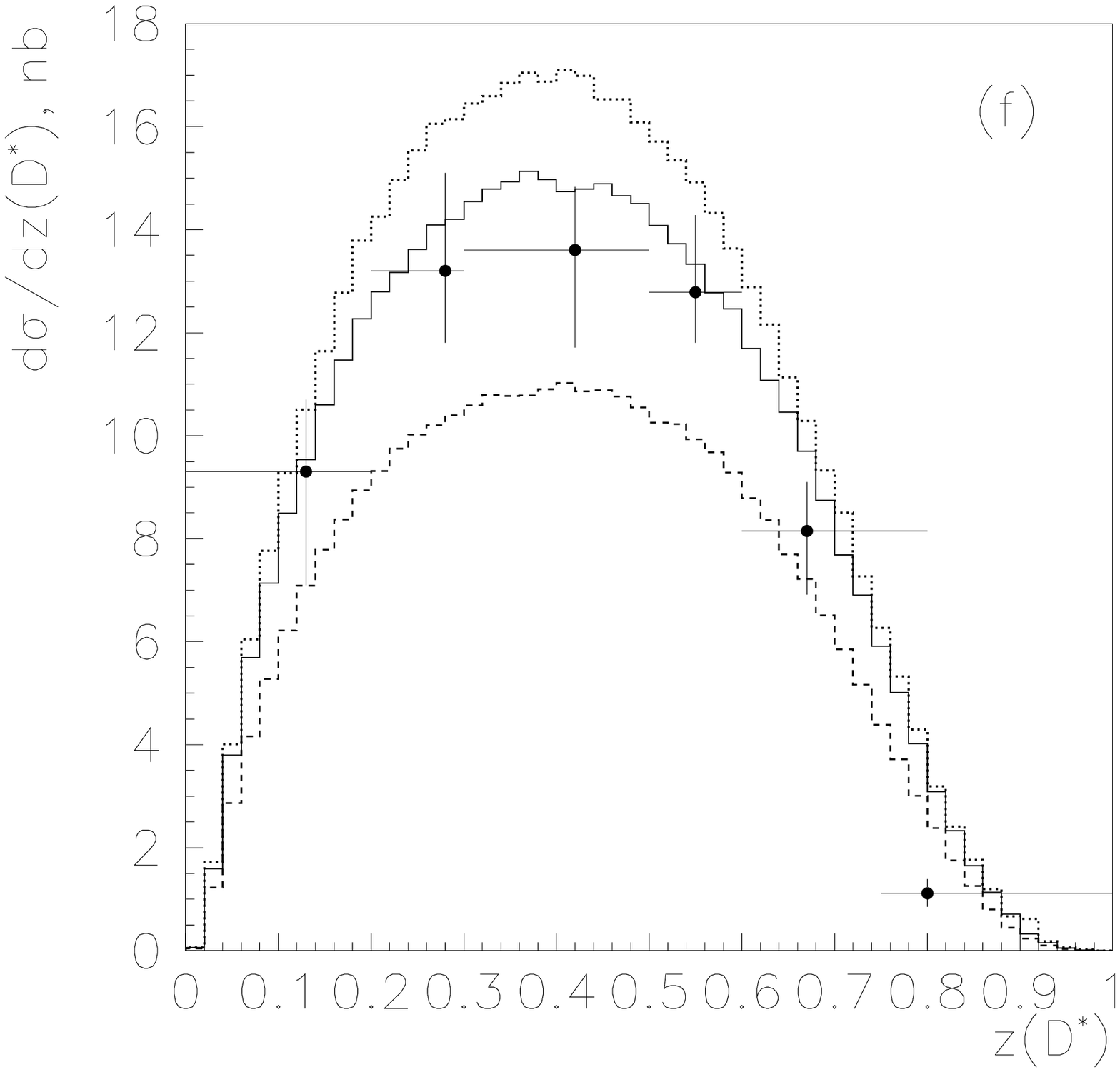}
\end{center}
\vspace*{-5mm}
\caption{Differential cross sections for deep inelastic $D^{*\pm}$ production in
the ZEUS accessible kinematical region as functions of:
$(a)\;\log_{10}Q^2$,~
$(b)\;\log_{10}x$,~
$(c)\;W$,~
$(d)\;p_T(D^*)$,
$(e)\;\eta(D^*)$ and
$(f)\;z(D^*)$.
 \label{fig:fig1}}
\end{figure*}

It is known that the resummation \cite{4,5,8,9} of the terms
$\;[\mbox{ln}(\mu^2/\Lambda^2)\,\alpha_s]^n,\;$
$[\mbox{ln}(1/x)\,\alpha_s]^n\;$ and  
$[\mbox{ln}(\mu^2/\Lambda^2)\,\mbox{ln}(1/x)\,\alpha_s]^n$
in SHA results in the so called unintegrated gluon distribution \\
${\cal F}(x,k_{T}^2, Q^2_0)$, which determines the probability 
to find a gluon  carrying the longitudinal momentum fraction $x$
and transverse momentum $k_T$ at the probing scale $Q^2_0$.
It obeys the BFKL equation~\cite{10} and reduces to the conventional gluon
density $G(x,\mu^2)$ once the $q_{T}$ dependence is integrated out:
\begin{equation} \label{kt}
\int_0^{\mu^2}\!\!{\cal F}(x,k_{T}^2,\,Q^2_0)\;dk_{T}^2{=}
x\,G(x,\mu^2).  
\end{equation}
The factorization  scale $Q^2_0$ (such that $\alpha_s(Q^2_0)$ ${<}1$) 
indicates the scale of the nonperturbative input distribution.

The CCFM evolution equation \cite{11} includes coherence effects via parton 
angular ordering and reproduces the BFKL evolution equation in the small
$x$ limit.
Therefore the CCFM unintegrated parton distribution 
${\cal A}(x,k_T^2,Q^2_0,\bar{q}^2)$ (unlike the function
${\cal F}(x,k_T^2,Q^2_0)$)
depends also on the maximum angle allowed for any emission 
corresponding to $\bar{q}=\vec{p}_t/(1-z)$.
In the small $x$ limit it reduces to ${\cal F}$~\cite{11}.

When calculating the spin average of the matrix element squared,
we substitute the full lepton tensor for the photon polarization matrix:
\begin{equation}
\overline{\epsilon_{\g}^{\mu}\epsilon_{\g}^{*\nu}}=
 [8p_e^{\mu}p_e^{\nu}-4(p_eq)g^{\mu\nu}]/(q^2)^2 \label{epsq}
\end{equation}
(including also the photon propagator factor).
The virtual gluon polarization matrix is taken in the form \cite{4}:
\begin{equation}
 \overline{\epsilon_g^{\mu}\epsilon_g^{*\nu}}
  =p_p^{\mu}p_p^{\nu}x^2/|k_{T}|^2=k_{T}^\mu k_{T}^\nu/|k_{T}|^2, \label{epsk}
\end{equation}
where $p_e$ and $p_p$ are the 4-momenta of the incoming electron and proton.

To parametrise the unintegrated structure functions, we use the prescriptions
of ref. \cite{12}. The proposed method lies upon a strai\- ghtforward
perturbative solution of the BFKL equation where the collinear gluon
density
$x\,G(x,\mu^2)$ is used as the boundary condition in the integral form (1).
Technically, the unintegrated gluon density is calculated as a convolution of
the collinear gluon density with universal weight factors:
\begin{equation}
{\cal F}(x,k_{t}^2,\mu^2)=
\int_x^1 {\cal G}(\eta,k_{t}^2,\mu^2)\, 
  \frac{x}{\eta}\,G(\frac{x}{\eta},\mu^2)\,d\eta,
\end{equation}
\begin{eqnarray} 
&&\hspace*{-7mm}{\cal G}(\eta,k_{t}^2,\mu^2)=
 \frac{\bar{\alpha}_s}{\eta\,k_{t}^2}\,
 J_0(2\sqrt{\bar{\alpha}_s\ln(1/\eta)\ln(\mu^2/k_{t}^2)}), \nonumber \\
&&k_{t}^2<\mu^2,
\end{eqnarray}
\begin{eqnarray}
&&\hspace*{-7mm}{\cal G}(\eta,k_{t}^2,\mu^2)=
  \frac{\bar{\alpha}_s}{\eta\,k_{t}^2}\,
 I_0(2\sqrt{\bar{\alpha}_s\ln(1/\eta)\ln(k_{t}^2/\mu^2)}), \nonumber \\
&&k_{t}^2>\mu^2,
\end{eqnarray}
where $J_0$ and $I_0$ stand for Bessel functions (of real and imaginary
arguments, respectively), and $\bar{\alpha}_s=\alpha_s/3\pi$.
The latter parameter is connected with the Pomeron trajectory intercept:
$\Delta=\bar{\alpha}_s\,4\ln{2}$ in the LO, and
$\Delta=\bar{\alpha}_s\,4\ln{2}-N\bar{\alpha}_s^2$ in the NLO approximations,
respectively, where $N \sim 18$ \cite{13}.

In the previous work \cite{6} we used the standard GRV parametrization
\cite{14} for the colline\-ar gluon density, from which the unintegrated
gluon distribution was developed according to eqs. (4) - (6).
Some other essential parameters were chosen as follows: the charm quark mass
$m_c=1.5$~GeV, the Peterson fragmentation parameter $\epsilon=0.06$, the
overall $c\to D^*$ fragmentation probability 0.26.
The Pomeron intercept $\Delta$ was regarded as free parameter, and then the
value $\Delta=0.35$ has been extracted from a fit to the experimental
$p_T(D^*)$ spectrum measured by the ZEUS collaboration \cite{15}.

 In this paper we also use an unintegrated gluon density coming from
a solution of the CCFM evolution equation (see \cite{16}). It was shown in \cite{16}
that a good description of the inclusive
structure function $F_2(x,Q^2)$ and the production of forward jets in DIS,
which are believed to be a prominent signature of small $x$ parton dynamics,
can be obtained using the CCFM unintegrated gluon distribution. In \cite{7}
we have used the hadron level Monte Carlo program CASCADE described in \cite{16}
to predict the cross section for $D*$ photoproduction at HERA energies.
It was shown that in this case the description of the differential cross section 
$d\sigma/d\eta(D^*)$ for $Q^2 < $ 1 GeV$^2$ for different regions of
$p_T(D^*)$ is improved.\\

\section{NUMERICAL RESULTS}

 The theoretical predictions on the differential cross sections of
deep inelastic $D^*$ production are shown in Figure~\ref{fig:fig1} for the ZEUS kinematical
region: $1\,<\, Q^2\, <\, 600$ GeV$^2$, \, $1.5\, <\, p_T(D^{*\pm})\, <\, 15$
GeV and $\eta (D^{*\pm})\, < \, 1.5$. Different curves in Figure~\ref{fig:fig1}
correspond  to different values of the Pomeron intercept parameter:
$\Delta$ = 0.166 (dotted), 0.35 (solid) and 0.53 (dashed).
 We see that the theoretical curves with $\Delta = 0.35$
describe all ZEUS experimental data \cite{3} on the differential cross
sections, except the $d\sigma/d\eta(D^*)$ distribution.
The SHA calculations with BFKL unintegrated gluon distribution show some 
shift to negative $\eta(D^*)$ with respect to the data.
This descrepancy between the data and the SHA prediction could result
from the use of the Peterson fragmentation function. Another reason
may be connected with the BFKL LO unitegrated gluon density.
\begin{figure}
\includegraphics[width=\linewidth]{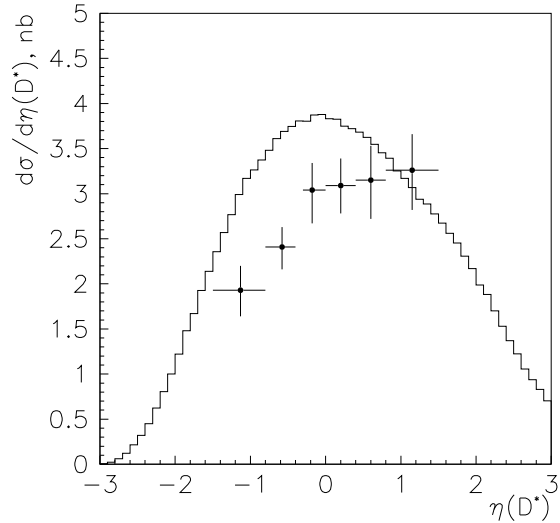}
\vspace*{-5mm}
\caption{$d\sigma/d\eta(D^*)$,
CCFM scheme with Peterson fragmentation.\label{fig:fig2}}
\end{figure} 
\begin{figure}
\includegraphics[width=\linewidth]{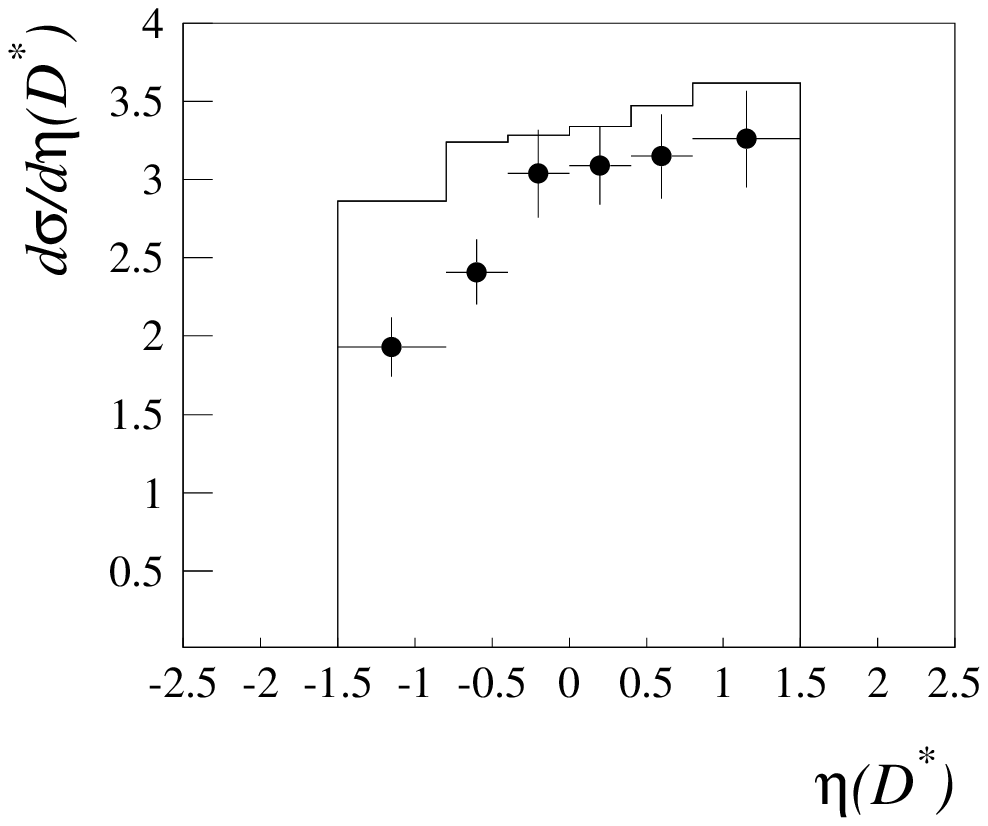}
\vspace*{-10mm}
\caption{$d\sigma/d\eta(D^*)$,
CCFM sheme with JETSET fragmentation.\label{fig:fig3}}
\end{figure}

The use of CCFM unintegrated gluon density (with angular ordering)
from MC generator CASCADE \cite{16} with the Peterson fragmentation function
for $c\to D^{*+}$ transition gives some positive shift to the
$d\sigma/d\eta(D^*)$ distribution (Figure~\ref{fig:fig2}) compared to BFKL curve
(see Figure~\ref{fig:fig1}e).
%
%
Finally, when we use the CCFM unintegrated gluon distribution
with more realistic JETSET based fragmentation function~\cite{JETSET}
implemented in \cite{16} we obtain  
good agreement between our theoretical results and the ZEUS experimental
data \cite{3} for $d\sigma/d\eta(D^*)$ (Figure~\ref{fig:fig3}).\\
\section{ACKNOWLEDGEMENTS}
One of us (N.Z.) would like to thank  Organizing Committee of the
Workshop "Diffraction 2000" for the invitation and financial support
and R. Fiore for exiting scientific atmosphere. This talk is
based on the work supported in part by the Royal Swedish Acad\-emy
of Sciences.

\end{document}